\documentclass[aps,noshowpacs,nobalancelastpage,floatfix,twocolumn,pra]{revtex4}
\usepackage{graphicx}
\usepackage{amsmath, amssymb}

\begin{document}

\title{Ultra-low-loss tapered optical fibers with minimal lengths}

\author{Ryutaro Nagai and Takao Aoki}

\affiliation{Department of Applied Physics, Waseda University, Tokyo, Japan}

\begin{abstract}
We design and fabricate ultra-low-loss tapered optical fibers (TOFs) with minimal lengths.
We first optimize variations of the torch scan length using the flame-brush method for fabricating TOFs with taper angles that satisfy the adiabaticity criteria. 
We accordingly fabricate TOFs with optimal shapes and compare their transmission to TOFs with a constant taper angle and TOFs with an exponential shape. 
The highest transmission measured for TOFs with an optimal shape is in excess of $99.7 \, \%$ with a total TOF length of only $23 \,{\rm mm}$, whereas TOFs with a constant taper angle of $2 \,{\rm mrad}$ reach $99.6 \, \%$ transmission for a $63 \,{\rm mm}$ TOF length.
\end{abstract}

\maketitle

\section{Introduction}

Tapered optical fibers with a sub-wavelength waist radius have been utilized for diverse applications in photonics and quantum optics \cite{Brambilla2010, Morrissey2013}. 
Strong confinement of the guided mode to a sub-wavelength radius leads to efficient nonlinear effects such as supercontinuum generation \cite{Birks2000, Leon-Saval2004, Gattass2006} and third-harmonic generation \cite{Akimov2003, Grubsky2005, Lee2012}. 
Large evanescent fields of the guided mode are beneficial for optical sensing and spectroscopy \cite{Villatoro2005, Polynkin2005, Zhang2008, Warken2007, Stiebeiner2009, Garcia-Fernandez2011}, for realizing strong light-matter interaction \cite{Klimov2004, Nayak2007, Vetsch2010, Goban2012, Fujiwara2011, Yalla2012, Kato2014, Chonan2014}, and for coupling to whispering-gallery-mode microresonators \cite{Knight1997, Cai2000}.

For these applications, minimizing TOF losses is important and sometimes crucial. 
In addition to extrinsic losses caused by surface roughness or contaminants, there are more fundamental TOF losses, 
caused by the power coupling from the fundamental mode to higher-order modes because of the non-zero taper angles.
These power coupling losses depend on the TOF shape; TOFs with smaller taper angles, {\it i.e.}, more adiabatic shapes, have lower power coupling losses. 
Transmissions of $99.4 \,\%$ and $99.95 \,\%$ have been achieved using TOFs with an exponential shape with a decay constant of $10 \,{\rm mm}$ \cite{Aoki2010} and using TOFs with a constant taper angle of $2 \,{\rm mrad}$ \cite{Ravets2013, Hoffman2014}, respectively. 
However, TOFs with small taper angles tend to be long (the lengths of the exponential TOF in Ref. 25 and the $2 \,{\rm mrad}$ TOF in Ref. 27 are $111 \,{\rm mm}$ and $84 \,{\rm mm}$, respectively);
there is a trade-off between the short length and high transmission of TOFs.

For some applications, {\it e.g.}, fiber coupling to microresonators, shorter TOFs are desirable because of their higher mechanical stability.
Moreover, there is sometimes a constraint on the length of the TOFs when they are installed in small spaces, such as cryostats \cite{Riviere2013} or vacuum chambers \cite{Aoki2006}.
Therefore, it is desirable to restrict the length of the TOFs as much as possible while maintaining high transmission. 
Useful guidelines for designing such TOFs are the adiabaticity criteria \cite{Love1986, Love1991, Birks1992}, which give the upper limit for the local taper angle as a function of the local taper radius. 
There have been studies reporting the fabrication of TOFs containing three tapered sections where each section has a different taper angle such that the taper angle is smaller where the adiabaticity criterion gives a smaller upper limit  \cite{Ravets2013, Hoffman2014, Stiebeiner2010}. 
Although transmissions higher than $99 \,\%$ have not been achieved with such trilinear TOFs, 
they can in principle have shorter TOF lengths than linear TOFs without sacrificing the transmission qualities.
However, there have been no reports on the design and fabrication of TOFs with the shapes optimized in systematic ways for achieving minimal TOF lengths.

In this paper, we design and fabricate TOFs that optimally satisfy the adiabaticity criterion. 
We adopt the flame-brush method \cite{Bilodeau1988} using variable torch scan lengths, 
and we optimize the scan length such that the local taper angle satisfies the adiabaticity criterion by taking a systematic approach.
We accordingly fabricate TOFs with optimal shapes and compare their transmission with those having a constant taper angle of $2 \,{\rm mrad}$ and those having an exponential shape with a decay constant of $3 \,{\rm mm}$. 
To investigate the reproducibility, five samples are fabricated for each profile.
The highest transmission measured for TOFs with an optimal shape reaches higher than $99.7 \,\%$ with a total TOF length of only $23 \,{\rm mm}$, whereas TOFs with a constant taper angle of $2 \,{\rm mrad}$ reach $99.6 \,\%$ transmission for a $63 \,{\rm mm}$ TOF length.

\section{Theory}

\subsection{Adiabaticity criteria}

A TOF is adiabatic if the taper angle is small enough such that there is negligible power coupling from the fundamental mode to higher-order modes.
Adiabaticity criteria have been proposed to delineate such conditions \cite{Love1986, Love1991, Birks1992}.
The delineation angle $\Omega(r)$ as a function of the local taper radius $r$ is given by the following length-scale criterion \cite{Love1991}:
\begin{equation}
 \Omega(r) = \frac{r}{2\pi}(\beta_{1}(r)-\beta_{2}(r)),
\end{equation}
where $\beta_{1}(r)$ and $\beta_{2}(r)$ are the propagation constants of the fundamental ($HE_{11}$) and the first excited ($HE_{12}$) modes, respectively. 
If the local taper angle $\theta(r)$ is much smaller than $\Omega(r)$, the mode coupling is negligible and the fundamental mode propagates adiabatically. 
We introduce an adiabaticity factor $F (>0)$ to express this condition as follows:
\begin{equation}
\theta(r) < F \Omega(r).
\label{eq:adiabatic}
\end{equation}
Smaller $F$ values lead to reduced mode coupling. 
Figure 1 shows the delineation angle of a step-index fiber calculated with Mathematica (Wolfram Research) using the three-layer model \cite{Karapetyan2012} 
with the wavelength $\lambda = 852 \,{\rm nm}$ and with the refractive indices and the radiuses of the core (cladding) of $n_{\rm core} = 1.4574$ ($n_{\rm clad}=1.4525$) and $r_{\rm core} = 2.4\, \mu {\rm m}$ ($r_{\rm clad} = 62.5 \,\mu {\rm m}$), respectively. 

\begin{figure}[htbp]
\includegraphics[width=8cm]{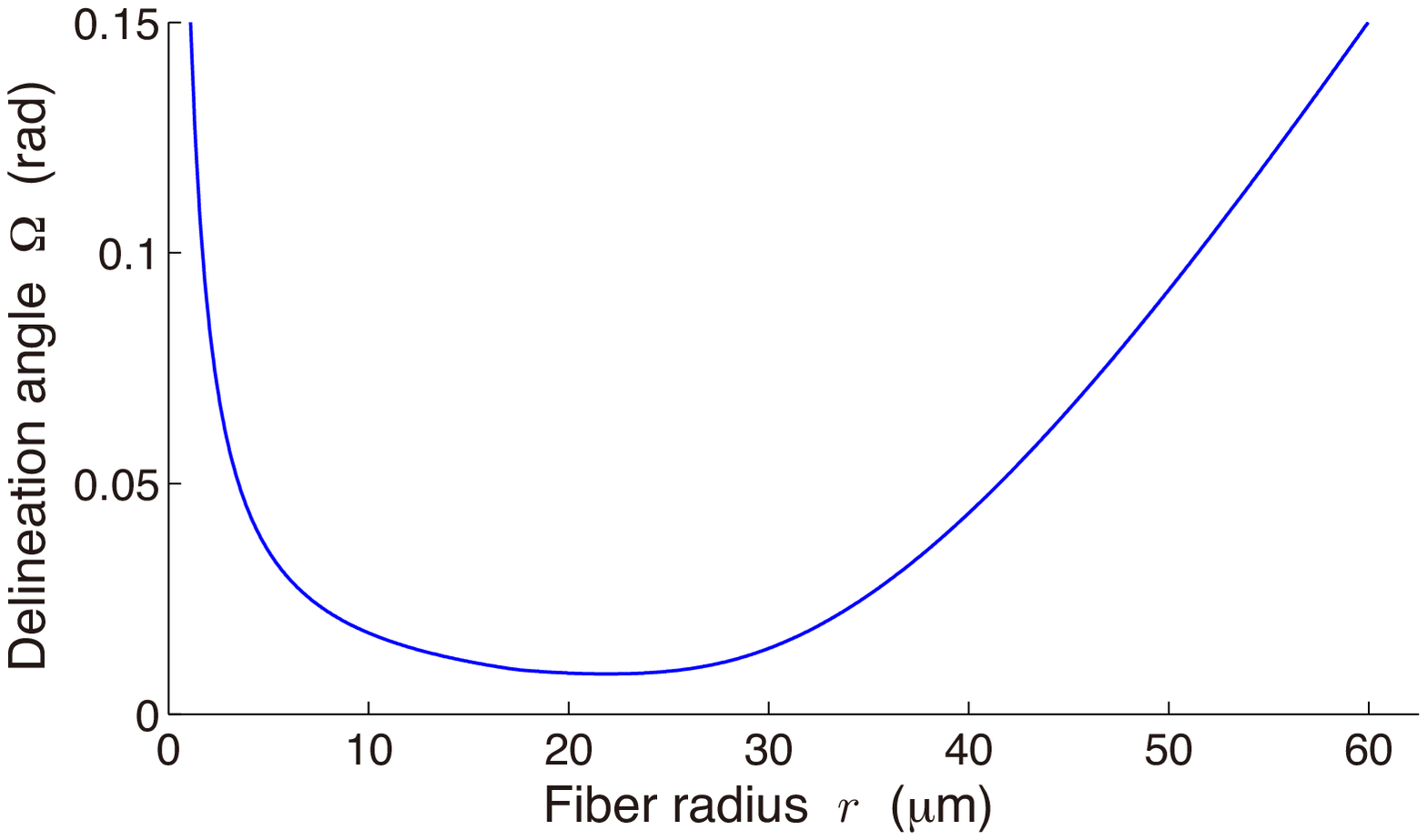}
\caption{The calculated delineation angle $\Omega$ as a function of the fiber radius $r$ of a step-index fiber for the wavelength $\lambda = 852 \, {\rm nm}$, where the refractive indices and the radii of the core (cladding) are $n_{\rm core} = 1.4574$ ($n_{\rm clad}=1.4525$) and $r_{\rm core} = 2.4\, \mu {\rm m}$ ($r_{\rm clad} = 62.5 \,\mu {\rm m}$), respectively. }
\end{figure}

\subsection{Model for the TOF shape} \label{sec:shape}

We model the shape of a TOF fabricated using the flame-brush method \cite{Bilodeau1988} with a variable torch scan length. 
In this method, a small flame heats a small region of a fiber with an initial radius $r_0$ to the softening point. 
While the fiber is pulled from both ends with a constant velocity $v_1$, the flame scans back and forth along the fiber axis with a constant velocity $v_2$. The length of the $k$-th scan is $L_k$.
Here we count $k$ by the number of one-way scans, not by the number of roundtrip scans.
This scanning process elongates the heated region of the fiber, and its radius is reduced to $r_k$ after the $k$-th scan of the torch.

\begin{figure}[htbp]
\includegraphics[width=8cm]{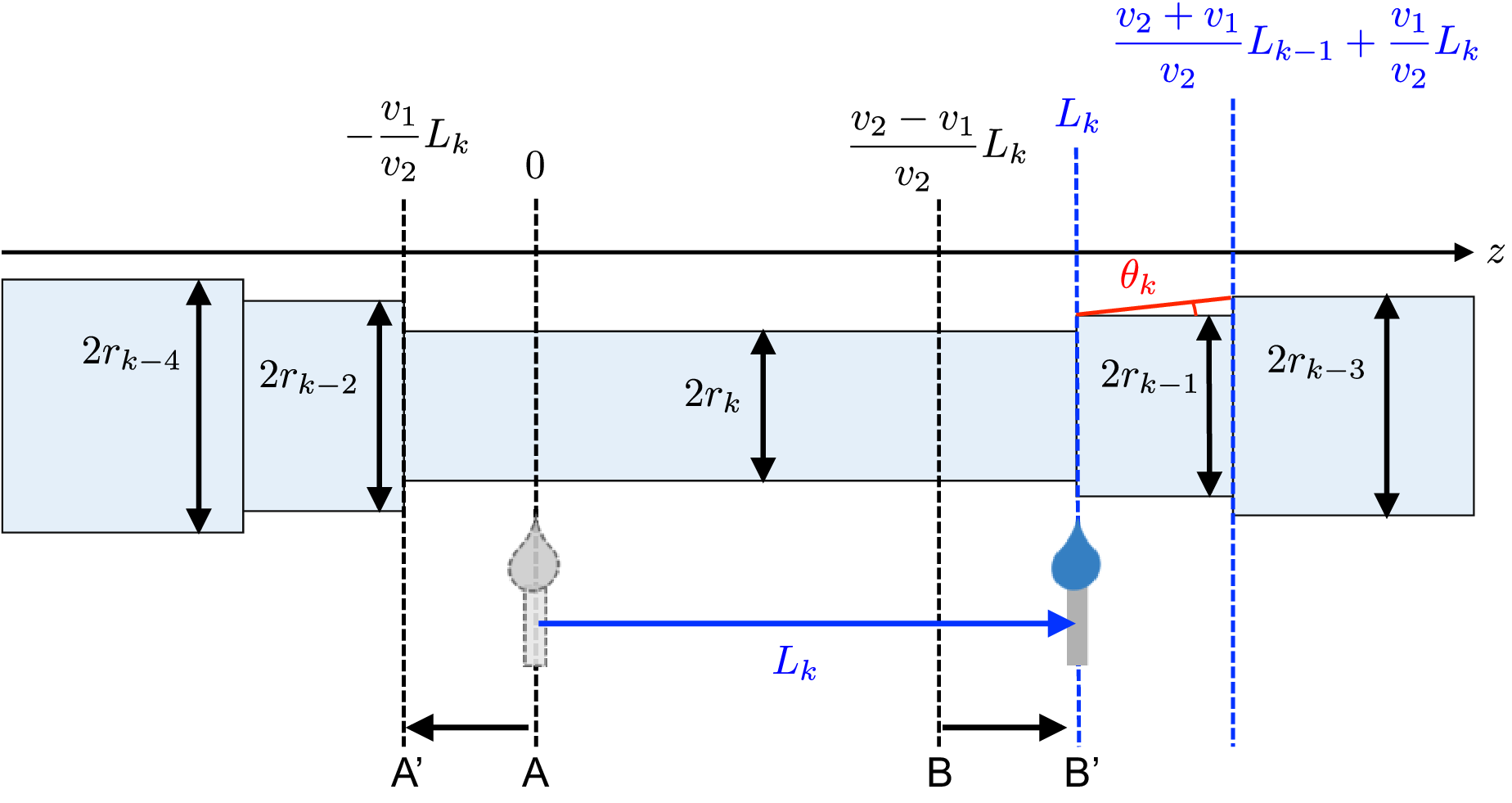}
\caption{Schematic of the TOF elongation after the $k$-th scan of the torch with the flame-brush method.
The origin of the $z$-axis is taken as the position of the torch at the beginning of the $k$-th scan 
({\it i.e.}, at the end of the $(k-1)$-th scan). 
The torch scans in the positive direction over a length of $L_k$, and the radius is reduced from $r_{k-1}$ to $r_k$ after the $k$-th scan.
}
\end{figure}

Figure 2 shows the schematic of the TOF after the $k$-th scan.
We assume that the flame width is small enough that a negligibly short length of the fiber is heated at any given time. 
As shown in Fig. 2, the cylinder with radius $r_{k-1}$ and length $\frac{v_2-v_1}{v_2}L_k$ between planes labeled A and B is heated during the $k$-th scan and become a cylinder with radius $r_{k}$ and length $\frac{v_2+v_1}{v_2}L_k$ between planes labeled A$^\prime$ and B$^\prime$ at the end of the $k$-th scan.
The conservation of the volume before and after the $k$-th scan,
\begin{equation}
\pi r_{k}^{2} \left(\frac{v_2+v_1}{v_2}\right)L_k = \pi r_{k-1}^{2} \left(\frac{v_2-v_1}{v_2}\right)L_k,
\end{equation}
leads to the a recurrence relation for the fiber radii:
\begin{equation}
r_{k}=\sqrt{\frac{v_{2}-v_{1}}{v_{2}+v_{1} } } r_{k-1}.
\end{equation}
The local taper angle $\theta_{k}$ after the $k$-th scan can be derived from Fig. 2 as 
\begin{equation}
\tan{\theta_{k}}=\frac{ r_{k-3}-r_{k-1} }{ \left( \frac{ v_{2} + v_{1} }{ v_{2} } L_{k-1} + \frac{v_{1} }{ v_{2} } L_{k} \right) - L_k} .
\end{equation}
Note that this model is invalid if the torch passes the end of the heated region for the $(k-2)$-th scan,
$L_k > \frac{ v_{2} + v_{1} }{ v_{2} } L_{k-1} + \frac{v_{1} }{ v_{2} } L_{k} $.
This requirement poses the limitation,
\begin{equation}
 L_{k} < \left( \frac{v_2+v_1}{v_2-v_1} \right) L_{k-1}  .
\label{eq:alpha}
\end{equation}
The total TOF length $L_{\rm total}$ after the $n$-th scan can be written as 
\begin{equation}
L_{\rm total} = \left( \frac{v_2 + v_1}{v_2} \right) L_{1} + \sum_{k=2}^{n}  \frac{2v_1}{v_2} L_{k}.
\end{equation}

\subsection{Optimization of the heat length profile}

To design an adiabatic TOF with minimal length, $L_{\it total}$ must be minimized under the condition $\theta_{k} < F\Omega(r=r_{k})$ for any $k$. 
For given values of $r_0$ and $v_1/v_2$, this minimization is an optimization of $L_{k}\ (k=1,2, \cdots, n)$.
Note that a smaller value of the ratio $v_1/v_2$ leads to smaller staircase-step changes in the fiber radius (see Eq. (4)), 
thus leading to a smoother taper at the cost of larger number of total scans $n$. 
In our typical experimental conditions, there are approximately 300 total scans, $n \sim 300$, which has too many degrees of freedom for a straightforward optimization. 
To reduce the degrees of freedom, we assume that $L_k$ is a smooth function of $k$, 
which should be a reasonable assumption, considering the smooth shape of $\Omega (r)$ (see Fig. 1).
Specifically, we consider a Lagrange interpolating polynomial, which passes through nine points that are fixed with even intervals on the scan-number axis $k$ and are variable on the $L$ axis, as $L_k$.

In addition to the conditions given by Eqs. (2) and (6), we pose the following conditions. 
First, we introduce the upper limit for $L_k$ of $L_{\rm upper} = 40\,000 \, \mu {\rm m}$ to take into account of the limitations of the experimental setup. 
Second, we introduce the lower limit for $L_k$ of $L_{\rm lower} = 400 \, \mu {\rm m}$ to take into account of the finite width of the flame. 
Furthermore, because the delineation angle $\Omega (r)$ abruptly increases as $r$ becomes less than $\sim 4 \,\mu {\rm m}$, 
we fix $L_k$ as $L_k = L_{\rm lower}$ for $r_k \lesssim 4 \, \mu {\rm m}$.
This restriction gives better results than optimizing $L_k$ for the whole indices of $k$.

We optimize $L_k$ with the downhill simplex method \cite{Press2007}, starting from ten sets of random initial values. 
The numerical calculation is performed with custom scripts using Matlab (MathWorks).
Figure 3 shows the optimized profiles of $L_{k}$ with the adiabaticity factors of $F=0.2$, 0.4, 0.6, and 0.8. 
The TOFs with larger $F$ values require longer scan lengths, which leads to longer TOF lengths.

\begin{figure}[htbp]
\includegraphics[width=8cm]{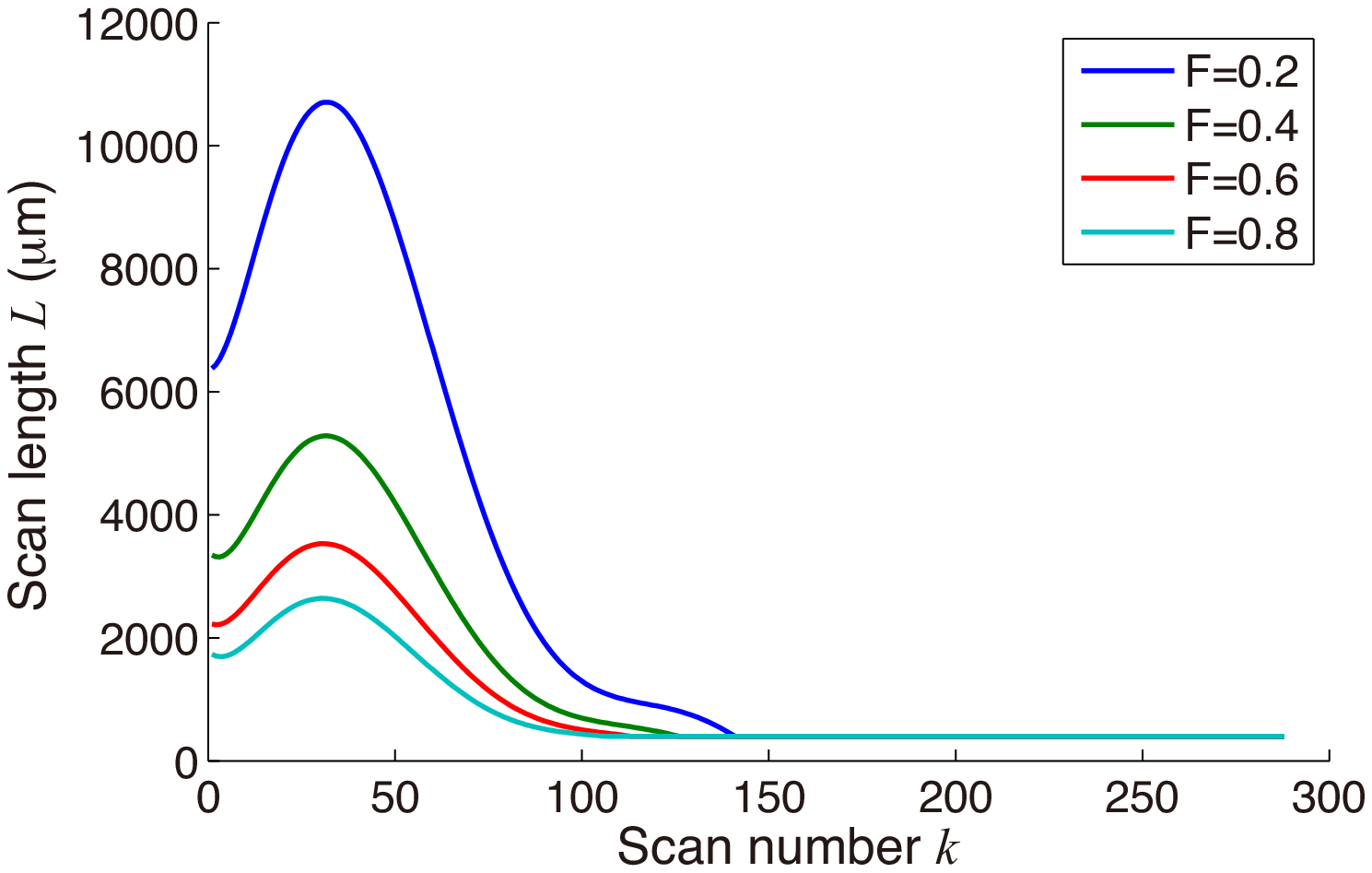}
\caption{Optimized profiles of $L_{k}$ for adiabaticity factors of $F=0.2$, 0.4, 0.6, and 0.8.}
\label{fig:profiles}
\end{figure}

\section{Experiment}

We experimentally fabricate TOFs according to the optimized $L_k$ profile obtained in the previous section. 
A commercial step-index single-mode fiber (Thorlabs, SM800) is fixed with two fiber clamps (Thorlabs, T711/M-250) on stepper-motor-driven linear translation stages (SURUGA SEIKI, KX1250), which pull the fiber on both ends at a velocity of $v_1 = 15 \, \mu {\rm m/s}$.
The fiber is heated with a flame from a hydrogen torch whose stainless-steel nozzle has a single hole with an inner diameter of $135 \,\mu {\rm m}$. 
We use pure hydrogen gas (no premix with oxygen), and the gas flow is adjusted to $10 \, {\rm mL/min}$ by a flow meter (KOFLOC, RK1250). The torch is also mounted on a stepper-motor-driven stage and scans along the fiber with a velocity of $v_2 = 750 \,\mu {\rm m/s}$.
Light from an $852 \, {\rm nm}$ laser diode is coupled to a 99/1-fiber beam splitter, and the $99\, \%$ output port is fusion spliced to the fiber being pulled.
We stabilize the laser output power to within $\pm 0.1\, \%$ by monitoring the power from the $1\, \%$ port and feeding it back to the diode current.
The TOF transmission is measured throughout the pull.

We fabricate TOFs optimized for the adiabaticity factors of $F=0.2, 0.4, 0.6,$ and 0.8. 
For comparison, we also fabricate TOFs with two other shapes.
One TOF has a constant taper angle of $2 \,{\rm mrad}$ for $r > 4\, \mu {\rm m}$ and an exponential shape with a waist length of $400 \,\mu {\rm m}$ for $r < 4\, \mu {\rm m}$.
Note that all of the optimized TOFs also have an exponential shape with a waist length of $400 \,\mu {\rm m}$ for $r \lesssim 4\, \mu {\rm m}$ because we fix $L_k$ as $L_k = 400 \, \mu {\rm m}$ for $r_k \lesssim 4 \, \mu {\rm m}$.
The other TOF has an exponential shape with a waist length of $3 \,{\rm mm}$.
Five samples are fabricated for each profile.
All samples are tapered to a waist radius that is thin enough to be single-mode guided ($r \lesssim 300 \,{\rm nm}$).

\begin{figure}[htbp]
\includegraphics[width=8cm]{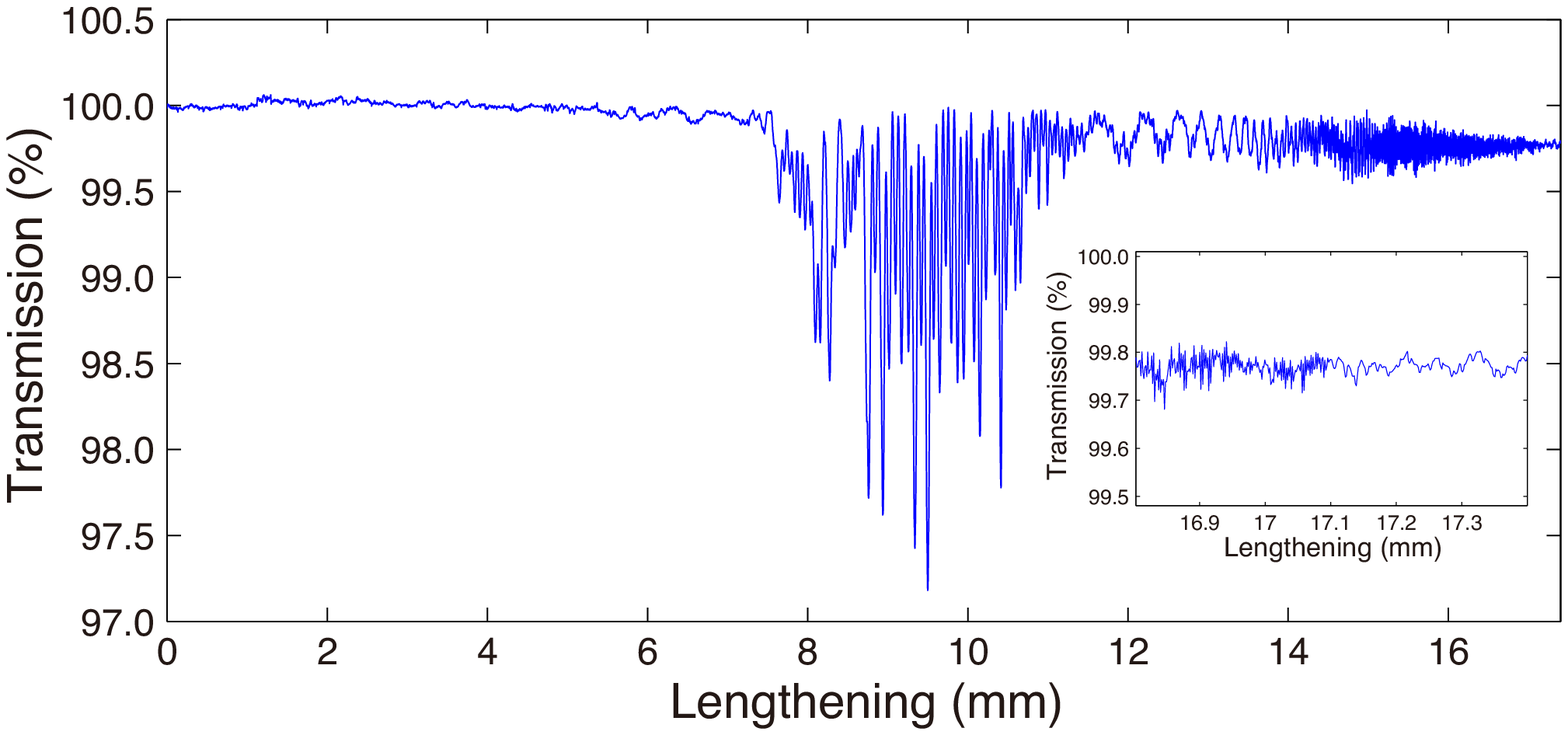}
\caption{The transmission change during the pull of the tapered fiber optimized with $F=0.4$.
The inset shows the magnified plot for the last $\sim 0.6 \, {\rm mm}$ of the pull.
}
\label{fig:F0.4}
\end{figure}

Figure 4 shows the change of the transmission during the pull for a TOF with $F=0.4$. 
The amplitude of the oscillation due to the coupling to higher-order modes is small, 
and the final transmission is in excess of $99.7 \, \%$.
The disappearance of the oscillation is clearly observed at the end of the pull, which is also confirmed with the short-time Fourier transform (spectrogram) of the transmission trace \cite{Orucevic2007}. 
The disappearance of the oscillation indicates the single-mode guidance.

\begin{figure}[htbp]
\includegraphics[width=8cm]{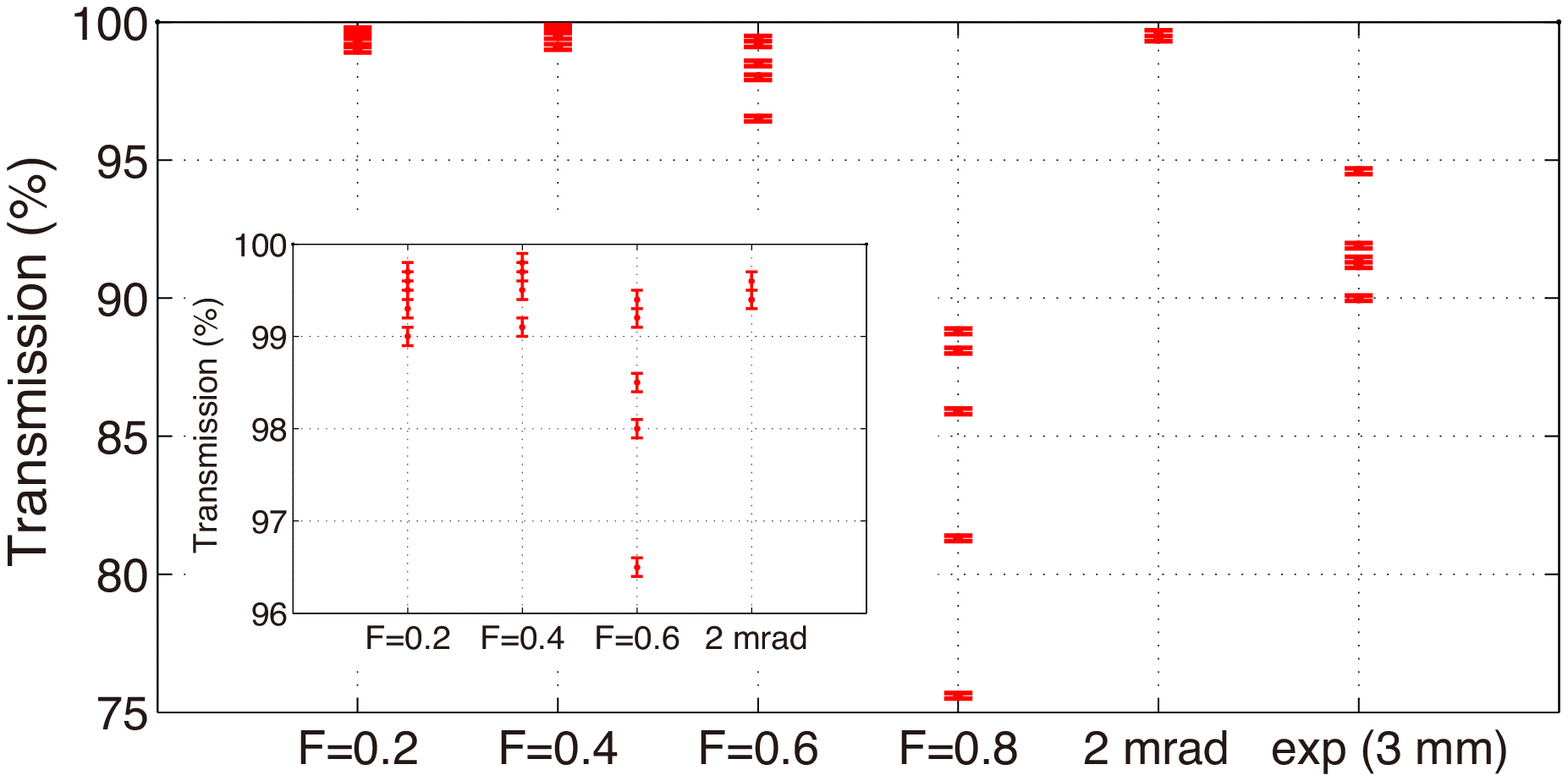}
\caption{The transmissions of fabricated TOFs with the adiabaticity factors of $F=0.2, 0.4, 0.6,$ and 0.8, a constant taper angle of $2 \,{\rm mrad}$, and an exponential shape with a decay constant of $3 \,{\rm mm}$. Five TOFs are fabricated for each profile. The inset shows the transmissions for $F=0.2, 0.4,$ and $2 \,{\rm mrad}$ with a magnified scale for the vertical axis.
The relative error in the transmission is $\pm 0.1\, \%$, which is limited by the laser output power stability.
}
\label{fig:T}
\end{figure}

Figure 5 summarizes the transmissions of all the fabricated TOFs.
There is a tendency that the profiles optimized with larger $F$ values tend to have larger losses.
Note that there are no noticeable differences in transmissions among the profiles of $F=0.2$, $F=0.4$, and the $2 \,{\rm mrad}$, in spite of the different taper angle profiles. 
This suggests that for these TOFs the losses due to mode coupling are sufficiently suppressed, and other extrinsic losses are dominant. 
Indeed, Ravets {\it et al.} reported the transmissions of $99.95 \,\%$ for $2 \,{\rm mrad}$ TOFs \cite{Ravets2013, Hoffman2014};
the highest transmission measured in our experiment for $2 \,{\rm mrad}$ TOFs is only $99.6 \,\%$. 
We suspect that the extrinsic losses for our TOFs are primarily caused by contaminants on the TOF surface 
because we use a homemade clean hood of class $100\,000$, compared to the class 100 cleanroom used in \cite{Ravets2013, Hoffman2014}.
Furthermore, perturbations such as air, mechanical vibrations of the setup, and fluctuation of the gas flow may cause deviation of the TOF shape from the theoretical prediction, leading to less adiabaticity and larger losses. 

\begin{figure}[htbp]
\includegraphics[width=8cm]{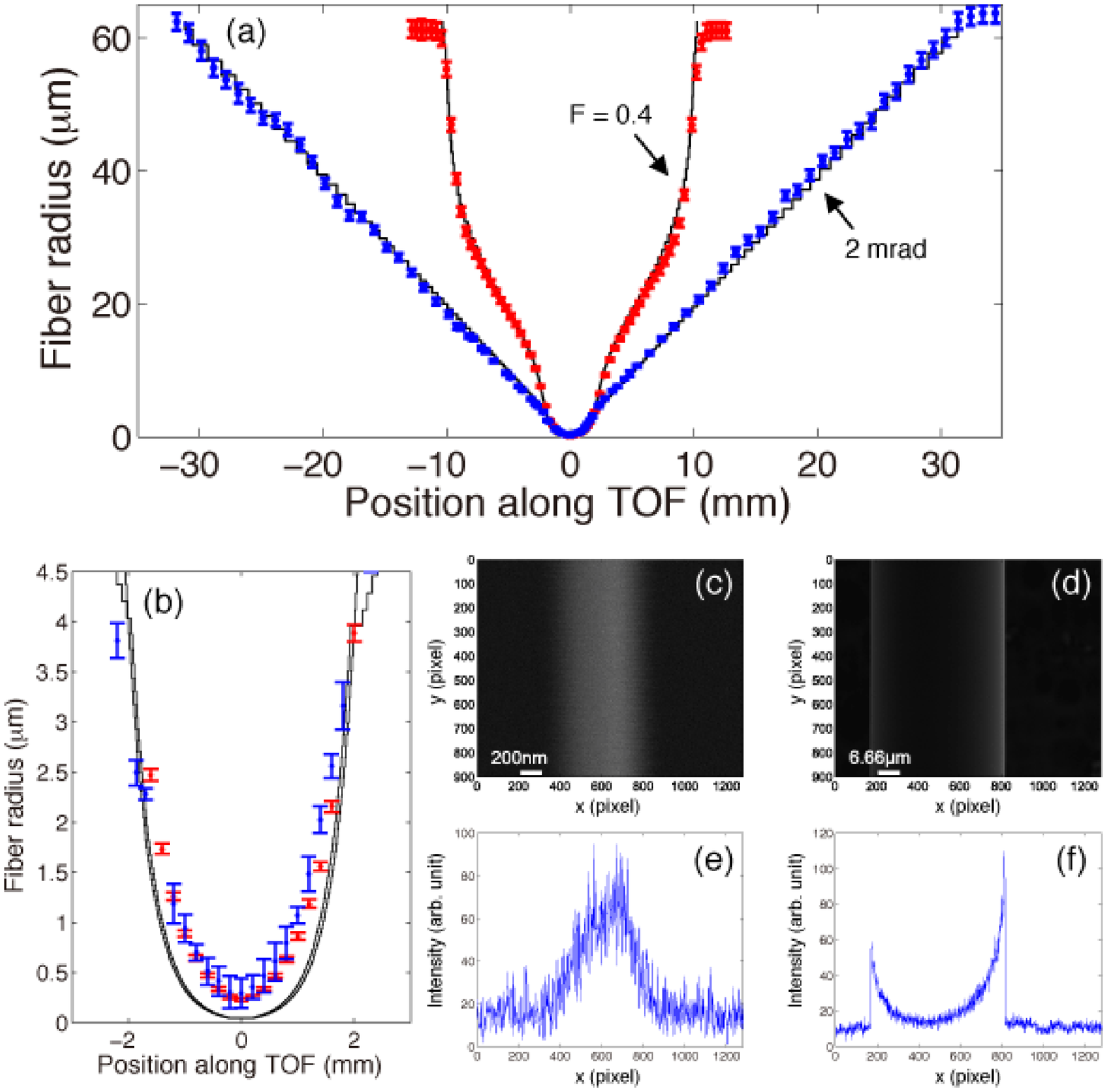}
\caption{(a) Comparison of the predicted and measured TOF shapes. 
Red and blue colored plots represent the shapes optimized for the $F=0.4$ and the $2 \,{\rm mrad}$ profiles, respectively. 
The solid lines are the theoretically predicted curves.
(b) The magnified plot for the region around the waist ($r \lesssim 4 \,\mu {\rm m}$) of (a).
Examples of SEM images of the $2 \,{\rm mrad}$ TOF (c) around the waist and (d) around $r \approx 20 \,\mu {\rm m}$, respectively.
(e) and (f) show the intensity profiles of (c) and (d), respectively.
}
\label{fig:compare}
\end{figure}

We measure the shape of the fabricated TOFs using a scanning electron microscope (SEM). 
Figure 6(a) and (b) show the measured shapes for the $F=0.4$ and $2 \,{\rm mrad}$ TOFs.
In practice, the TOFs have to be pulled over a longer distance than that predicted by the above model to achieve single-mode guidance at the waist. 
Longer pulling is required because of the effect of the finite flame width that is not negligible compared to the relatively small scan length $L_{\rm lower}$. 
Thus, we extend the scan length profiles $L_{k}$ to a larger scan number until single-mode guidance is reached, 
which is confirmed by the observed disappearance of the oscillation in the transmission (see Fig. 4). 
The predicted shapes in Fig. 6 are calculated with experimentally executed scan numbers. 
The TOF radii are measured as the half-width at the half-maximum of the each fiber edge slope in the intensity profile of the SEM images, as shown in Fig. 6(c)-(f).
The upper (lower) error bars are determined as the half-width at $10 \,\%$ ($90 \,\%$) of the maximum intensity of each slope. 
The SEM images at around the waist of TOFs ($r \lesssim 4 \,\mu {\rm m}$) are blurry due to the mechanical vibrations of the TOF, 
and the measured radii inevitably have large errors and may be overestimated (Fig. 6(c) and (e)), compared to those at $r \gtrsim 4 \,\mu {\rm m}$ (Fig. 6(d) and (f)). 
However, because the delineation angle $\Omega (r)$ abruptly increases at $r \lesssim 4 \,\mu {\rm m}$, precise knowledge of the TOF radius at these regions is not crucial to our study. 
For the regions with $r \gtrsim 4 \,\mu {\rm m}$, the shapes of the fabricated TOFs agree well with those predicted by the model 
with a normalized residual of $\lesssim 10 \,\%$.
Moreover, in spite of the higher transmission, the $F=0.4$ TOF has almost three times shorter length ($23 \,{\rm mm}$) than that for the $2 \,{\rm mrad}$ TOF ($63 \,{\rm mm}$).
Note that the error bars around the waist of the $F=0.4$ TOF are smaller than that for the $2 \,{\rm mrad}$ TOF 
because $F=0.4$ TOF has smaller mechanical vibrations due to its shorter TOF length.

\begin{figure}[htbp]
\includegraphics[width=8cm]{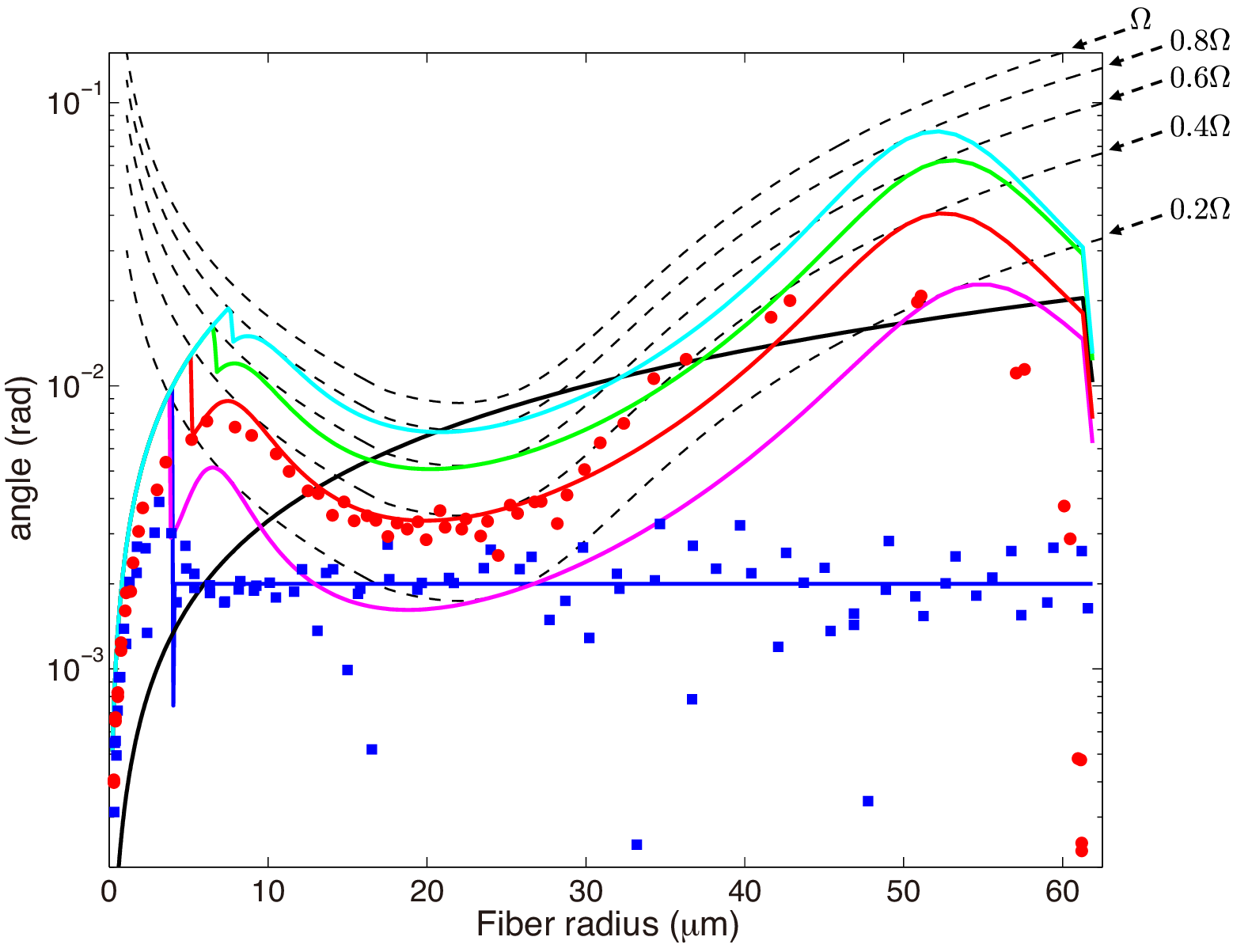}
\caption{
Comparison of the predicted and measured taper angles.
The cyan, green, red, magenta, blue, and black solid lines are the predicted taper angles for $F=0.8, 0.6, 0.4, 0.2$, $2 \,{\rm mrad}$, and 3 mm exponential TOFs, respectively. Red circles and blue squares are measured taper angles for the fabricated $F=0.4$ and $2 \,{\rm mrad}$ TOFs, respectively. The black dashed lines are $F\Omega$ with $F=1.0, 0.8, 0.6, 0.4, 0.2$.
}
\label{fig:compare}
\end{figure}

Figure 7 shows the taper angles as a function of TOF radius for the fabricated $F=0.4$ TOF and $2 \,{\rm mrad}$ TOF, along with the theoretically predicted taper angles for the optimized shape with $F=0.8, 0.6, 0.4, 0.2$, $2 \,{\rm mrad}$, and exponential TOFs. 
The measured taper angles of the $F=0.4$ TOF exceed the theoretical angles at around $30 \lesssim r \lesssim 40 \,\mu {\rm m}$. 
However, because the theoretical angles for the optimized profile coincidentally have much lower values than the targeted angles ($0.4 \Omega$) in this region, the measured angles do not considerably exceed the targeted angles.

\section{Conclusion}
In conclusion, we have designed and fabricated TOFs with high transmissions and minimal lengths. 
Transmission in excess of $99.7 \,\%$ has been observed for a TOF with a total length of $23 \,{\rm mm}$. 
We find that the losses due to the power coupling to higher-order modes are sufficiently suppressed, 
and the losses are dominated by other extrinsic losses, such as contaminants on the TOF surface. 
Further improvement of the optimization may be possible by increasing the degrees of freedom for the approximation of $L_k$ at the cost of longer computation time.
This work will enable the production of compact, stable setups for coupling whispering-gallery-mode microresonators to TOFs without sacrificing the transmission quality. 
We plan to build such a setup for the study of cavity quantum electrodynamics with laser-cooled atoms and microtoroidal resonators\cite{Aoki2006}.

\section*{Acknowledgment}
This work is partially supported by MATSUO FOUNDATION, JSPS KAKENHI Grant Number 26707022, Waseda University Grant for Special Research Projects (Project number: 2013A-501), and a research grant from The Murata Science Foundation.

\end{document}